\documentstyle[prl,aps,multicol,graphicx]{revtex}

\newcommand{\ev}[1]{\mbox{\bf E}\left(#1\right)}

\newcommand{\be}{\begin{equation}}
\newcommand{\ee}{\end{equation}}
\newcommand{\bs}{\begin{mathletters}} 
\newcommand{\es}{\end{mathletters}} 

\newcommand{\baa}{\begin{eqnarray}}
\newcommand{\eaa}{\end{eqnarray}}

\newcommand{\ba}{\bs\begin{eqnarray}}
\newcommand{\ea}{\end{eqnarray}\es}

\newcommand{\bt}[1]{\bs\label{#1}\begin{eqnarray}}
\newcommand{\et}{\end{eqnarray}\es}

\newcommand{\paper}[6]{#1, #2 #3{\bf #4}, #5 (19#6)}
\newcommand{\paperMillenium}[6]{#1, #2 #3{\bf #4}, #5 (20#6)}

\newcommand{\reffig}[1]{Fig.~\ref{#1}}

\newlength{\www}

\begin{document}
\draft
\preprint{ULDF-TH-number/month/year}

\title{Adaptive Optimization of Wave Functions for\\
Fermion Lattice Models}

\author{Matteo Beccaria$^{a, b}$, Antonio Moro$^{a, b}$}

\address{
${}^a$ Dipartimento di Fisica dell'Universit\`a di Lecce, I-73100, Italy,\\
${}^b$ Istituto Nazionale di Fisica Nucleare - INFN, Sezione di Lecce}

\maketitle

\begin{abstract}
We present a simulation algorithm for Hamiltonian
fermion lattice models. A guiding trial wave function is adaptively 
optimized during Monte Carlo evolution. We apply the method to the
two dimensional Gross-Neveu model and analyze systematc errors in the study of 
ground state properties. We show that accurate measurements can be
achieved by a proper extrapolation in the algorithm 
free parameters.
\end{abstract}


\begin{multicols}{2}
\narrowtext

Lattice Field Theory is a constructive framework where non-perturbative
properties of quantum models can be addressed
both analytically and by
numerical techniques.
The main standing theoretical viewpoints are the traditional Lagrangian 
approach~\cite{Lagrangian} and 
the Hamiltonian formulation~\cite{Hamiltonian}.
In the study of fermionic models, Lagrangian simulations 
suffer the drawback of requiring Grassmann variables that are difficult to
handle numerically and must be integrated out explicitly leading to
large non-local determinants. 
Instead, in the Hamiltonian approach,
the treatment of Fermi anticommuting operators 
is straightforward. In particular, this holds in one
spatial dimension where notoriously difficult 
{\em sign-problems}~\cite{GSP} are tame.

Another important reason to resort to Hamiltonian methods is
that they rely on powerful well founded Many-Body 
techniques~\cite{Linden}. In particular,
a direct analysis of the ground state structure is often 
feasible through a guiding {\em trial wave function}~\cite{WaveFunctions}.
This is an approximation to the exact ground state that 
can provide deep physical insights about the model under consideration.
Also, it plays a central role in the 
simulation algorithms and the quality
of the results depends critically on its accuracy~\cite{IS}. 
Usually, it contains a set of free parameters that deserve
optimization by rather expensive 
variational calculations~\cite{optimization}.

Here, we present a Monte Carlo (MC) algorithm that includes
automatic optimization of the trial wave function by means
of a non-linear feedback between state sampling and guiding.
The MC core is based on a general stochastic representation of 
matrix evolution problems~\cite{Jona} and has been discussed in the 
specific case of the Hubbard model~\cite{BeccariaFermions}.
The adaptive optimization strategy has been already applied
to Diffusion MC studies of purely bosonic models with continuous state 
space~\cite{WaveFunctions}.

In this Report, we focus on fermionic models and present 
an algorithm suitable for the study of Hamiltonians acting on a 
finite-dimensional fully discrete state space.
In fact, for a local fermion model discretized on a finite lattice, 
the Hamiltonian is a large sparse matrix
$H=\{H_{ss'}\}_{s, s'\in S}$, with $S$ denoting the discrete state space. The 
ground state can be obtained by acting on a given initial state
with the evolution semigroup $\Omega = \{e^{-t H}\}_{t\ge
0}$ in the $t\to\infty$ limit. 
For simplicity, we assume a non degenerate ground state,
in the general case $\Omega$ projects onto the lowest eigenspace.

To build a MC algorithm, we need 
a probabilistic representation of $\Omega$. 
For each pair $s, s'\in S$ such that $s\neq s'$ and $H_{s's}\neq 0$
we define $\Gamma_{s's} = -H_{s's}$. We assume that all
$\Gamma_{s's}>0$ (no sign-problem)
and build a $S$-valued Markov stochastic process $s_t$
by identifying $\Gamma_{s's}$ as the rate for
the transition $s\to s'$. Hence, 
the average occupation 
$
P_s(t) = \ev{\delta_{s, s_t}},
$
with $\ev{\cdot}$ denoting the average with respect 
to $s_t$, obeys the Master Equation 
$
\dot P_s(
\beta) = \sum_{s'\neq s}(\Gamma_{ss'}
P_{s'}-\Gamma_{s's} P_s) .  
$

Related to $s_t$, we also define the real valued stochastic process 
$
W_t = \exp\left(-\int_0^t \omega_{s_t}\ dt\right) ,
$
with $\omega_s = \sum_{s'\in S} H_{s's}$.
It can be shown that 
the weighted expectation value $\psi_s(t) = \ev{\delta_{s, s_t} W_t}$
reconstructs $\Omega$:
$$
\frac{d}{dt}\psi_s(t) = -\sum_{s'\in S} H_{s s'}\psi_{s'}(t) ,
$$
with $\psi_s(0) = \mbox{Prob}(s_0=s)$. 
Matrix elements of $\Omega$
can be identified with certain expectation values. 
In particular, the ground state
energy $E_0$ can be obtained by 
\be
\label{energy}
E_0 = \lim_{t\to +\infty} \frac{\ev{\omega_{s_t}\ W_t}}{\ev{W_t}} ,
\ee
that gives $E_0$ as the asymptotic average of $\omega_s$ over 
realizations of $s_t$ with weight $W_t$, called {\em walkers} in the
following.
The actual construction of the process is straighforward.
A realization of $s_t$
 is a piece-wise constant map ${\bf R}\to S$
with isolated jumps at times $t=t_0, t_1, \dots$, with
$t_0<t_1<t_2<\cdots$. 
An algorithm to compute the triples $\{t_n, s_{t_n},
W_{t_n}\}$ is the following:
\begin{enumerate}

\item We simply denote $s_{t_n} \equiv s$ and 
define the set $T_s$ of target states connected to $s$:
$T_s = \{s', \Gamma_{s's}>0\}$. We also define the 
total width $\Gamma_s = \sum_{s'\in T_s} \Gamma_{s's}$.

\item Extract $\tau\ge 0$ with probability density 
$p_s(\tau) = \Gamma_s e^{-\Gamma_s\tau}$. 
In other words, $\tau = -\frac 1 {\Gamma_s} \log\xi$ with $\xi$ 
uniformly distributed in $[0,1]$. 

\item Extract a new state $s'\in T_s$ with probability
$p_{s'} = \Gamma_{s's}/\Gamma_s$.

\item Define $t_{n+1} = t_n + \tau$, $s_{t_{n+1}} = s'$ and 
$W_{t_{n+1}} = W_{t_n}\cdot e^{-\omega_s\tau}$.
\end{enumerate}
The above algorithm is the explicit zero imaginary time limit 
of power algorithms~\cite{SorellaSign}.

For a better performance, it is useful to introduce 
a trial state $|\Phi(\alpha)\rangle$ depending on some parameters
$\alpha$. 
The original Hamiltonian $H$ is replaced by the
isospectral 
$H_{ss'}(\alpha)= \Phi_{s}(\alpha) H_{ss'}\Phi_{s'}^{-1}(\alpha)$
with $\Phi_s(\alpha) = \langle s | \Phi(\alpha)\rangle$.
The algorithm is unchanged (hermiticity of $H(\alpha)$ has not 
been assumed), 
but everything, in particular $\omega_s$, becomes $\alpha$-dependent.
In the ideal case when 
$|\Phi(\alpha)\rangle$ is the exact ground state, then $\omega_s\equiv 
E_0$ and the ground state energy is estimated by Eq.~(\ref{energy}) with
zero fluctuations.

As is well known, a naive implementation of Eq.~(\ref{energy}) fails
because the variance of the right hand side diverges as
$t\to +\infty$. A possible way out is 
Stochastic
Reconfiguration (SR)~\cite{hether,SR,SorellaHeisenberg,SorellaSign}. 
An ensemble with a large fixed number $K$ of walkers is introduced and 
a branching procedure deletes
walkers with low weight and makes copies of 
the ones with larger weight. In the end, we take the numerical 
limit $K\to\infty$.
If $\beta$ is the time between two SR, then
we denote the estimate of the ground state energy by
$\widehat{E_0}(\beta, K, \alpha)$ where we do not write 
the dependence on physical parameters (lattice size, couplings).
Usually, the dependence on $\alpha$ is quite strong and requires
optimization to make $|\Phi(\alpha)\rangle$ the closest possible
to the exact ground state.

As we remarked, 
a possible way to optimize $\alpha$ is to minimize the fluctuations of
$\omega_{s_t}(\alpha)$~\cite{Wilson}. 
To this aim, following the general ideas of~\cite{Koch}, 
we promote $\alpha$ to a sequence $\{\alpha_n\}$ and
after each SR, 
we compute the variance of $\omega(\alpha)$ over the $K$ walkers,
with their states kept fixed.
Then, we propose to update $\alpha$ according to 
\be
\alpha_{n+1} = \alpha_n -\eta_n\nabla_{\alpha_n}
\mbox{Var}\ \omega(\alpha_n) .
\ee
The sequence $\{\eta_n\}$ controls the speed of the adaptive process
and vanishes as $n\to\infty$, typically like 
$n^{-1}$.
The novelty of the procedure is that MC
 sampling and trial wave function optimization are coupled. 
A change in $\alpha$ induces a change in the walker dynamical
distribution which in turn determines the next evolution of
$\alpha$. The whole process is non-linear and an explicit 
numerical investigation is required to assess its stability.

As a specific non-trivial application, we consider 
the two dimensional Gross-Neveu model~\cite{GrossNeveu1}
described by the Hamiltonian
\be
H = \int dx \left[
-i\psi^{a \dagger}\sigma_x\partial_x\psi^a-\frac{g^2}{2N_f}(\psi^{a
\dagger}\sigma_z \psi^a)^2 \right],
\ee
where $\psi^a$ are $N_f$ Dirac fermions and we sum over the repeated 
flavor index $a=1,\dots, N_f$.
The model is asymptotically free, admits a $1/N_f$ 
expansion and breaks spontaneously the 
discrete chiral $Z_2$ symmetry $\psi\to \gamma_5\psi$.

Following~\cite{GrossNeveu2}, a lattice formulation with 
staggered Kogut-Susskind fermions~\cite{KS} is based on
$$
H = -\sum_{n=0}^{L-1} \left\{
\frac 1 2 (c_n^{a \dagger} c_{n+1}^a + \mbox{h.c.}) + \frac{g^2}{8N_f}
(c_n^{a \dagger} c_n^a-c_{n+1}^{a \dagger} c_{n+1}^a)^2
\right\}
$$
where $\{c^a_n, c^b_m\}=0$, $\{c^a_n, c_m^{b \dagger}\}=\delta_{n,m}
\delta_{a,b}$ and
periodic boundary conditions are assumed. 
The state space is the set of eigenstates of the occupation number
operators $n_i^a = c_i^{a \dagger}c_i^a$ denoted by $|{\bf n}\rangle$.
The fermion number is conserved and we focus on the half-filled
sector with $\sum_i n_i^a = L/2$.
The $Z_2$ symmetry corresponds to translations 
by two lattice sites.
To avoid sign-problems related to boundary crossing 
we choose in the following $L\ \mbox{mod}\ 4 = 2$ (the ground state
is then non-degenerate).

We adopt the one parameter trial wave function 
$$
\langle {\bf n} | \Phi(\alpha)\rangle = 
\exp\left[\alpha\sum_{i=0}^{L-1} (\sum_{a=1}^{N_f} (n_i^a-n_{i+1}^a))^2\right]
\langle {\bf n} | g=0\rangle,
$$
where $|g=0\rangle$ is the exact ground state at $g=0$.
The algorithm requires an explicit formula for the ratio
$\langle {\bf n'}  |\Phi\rangle /\langle 
{\bf n}  |\Phi\rangle$ where $|{\bf n}\rangle$ and $|{\bf n'}\rangle$
are states that differ by one fermion hopping. If 
$\{x_i\}$ and $\{x_i'\}$ are the $L/2$ fermion positions in the two
states and if $x_i = x_i'$ for $i\neq p$, then the following 
formula can be derived
$$
\frac{\langle {\bf n'}  |\Phi\rangle}{\langle {\bf n}  |
\Phi\rangle} =
e^{\frac{2\pi i}{L}\frac{L/2-1}{2}(x_p-x_p')}
 \frac{
\prod_{k\neq p}\left(\exp\frac{2\pi i x_p'}{L}-\exp\frac{2\pi i
x_k}{L}\right)}
{
\prod_{k\neq p}\left(\exp\frac{2\pi i x_p}{L}-\exp\frac{2\pi i
x_k}{L}\right)} .
$$

We compute the ground state energy on a lattice with $L=10$ sites
and begin our analysis with the case $N_f=2$. We consider several
ensemble sizes and evolution times: $K=10$, $50$, $100$ and $500$,
$\beta = 0.1$, $0.25$, $0.5$ and $1.0$. For each pair $(K, \beta)$ we
determine by the adaptive algorithm the best $\alpha$ and estimate
the ground state energy. 
For comparison, we also determine $E_0$ by exact Lanczos diagonalization.

\begin{figure}[htb]
\centerline{\includegraphics*[width=4.75cm,angle=-90]{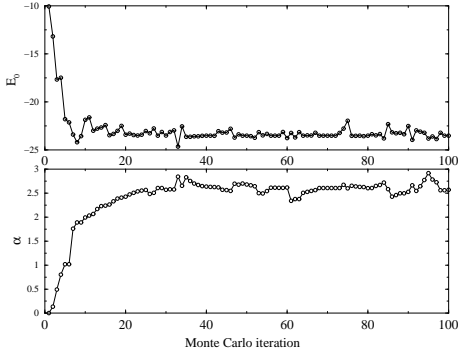}}
\caption{$L=10$, $N_f=2$, $g=3.0$, $K=10$, $\beta = 0.5$.
MC evolution of the ground state energy estimate and
of the $\alpha$ parameter.}
\label{fig:1}
\end{figure}

\begin{figure}[htb]
\centerline{\includegraphics*[width=4.75cm,angle=-90]{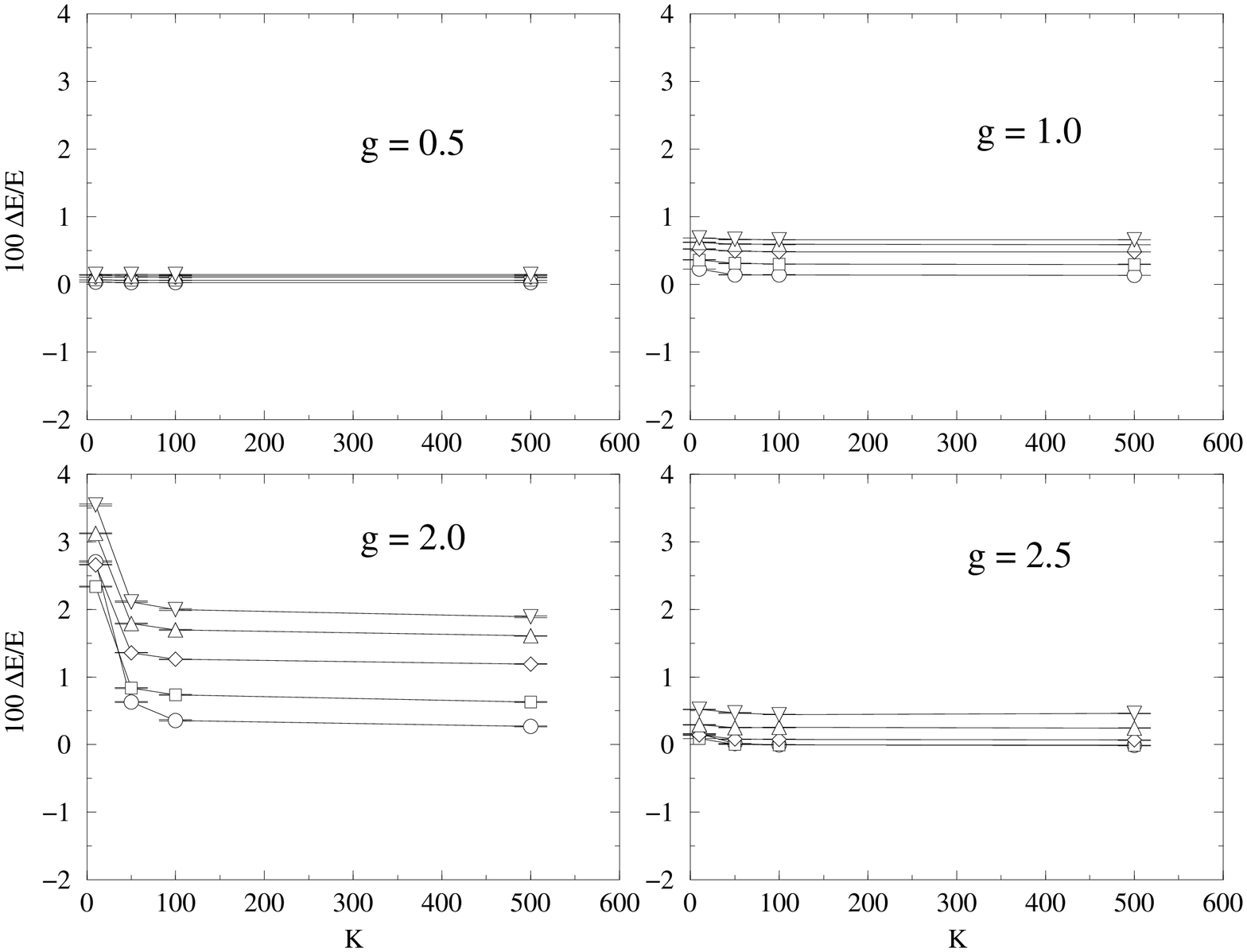}}
\caption{$L=10$, $N_f=2$. 
Relative percentual error on the ground state energy. 
The various lines correspond to 
$\beta = 0.1$ (circles), 
$\beta = 0.25$ (squares),
$\beta = 0.5$ (diamonds), 
$\beta = 0.75$ (triangles up) and 
$\beta = 1.0$ (triangles down).
}
\label{fig:2}
\end{figure}

\begin{figure}[h]
\centerline{\includegraphics*[width=4.75cm,angle=-90]{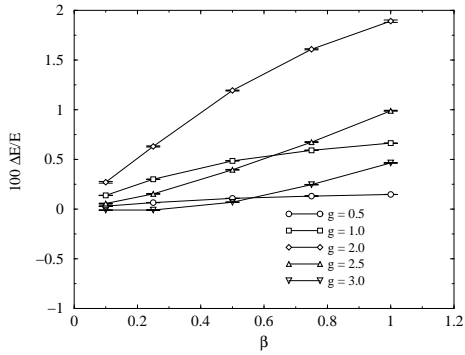}}
\caption{$L=10$, $N_f=2$, $K=500$. 
Relative percentual error on the energy obtained
from data at large $K$ at several $\beta$.}
\label{fig:3}
\end{figure}

\reffig{fig:1} shows the typical initial steps of a run.
The parameter $\alpha$ and the energy measurements evolve and 
fluctuate
around $(K, \beta)$ dependent definite average values 
$\alpha^*(K, \beta)$ and $\widehat{E_0}(K, \beta, \alpha^*(K, \beta))$. 
For large $K$, 
the statistical error on $\widehat{E_0}$ decreases like $K^{-1/2}$.
For $K\to \infty$, the results are expected to be $\beta$ independent. 
However, for moderate ensemble sizes, like those considered ($K\sim 500$),
a residual $\beta$ dependence can be observed, particularly at 
intermediate coupling, as shown in~\reffig{fig:2}. 
This effect is due to the process of walker selection associated to 
SR. The correct approach is to take the $\beta\to 0$ limit
where this effect is expected to be negligible.
In~\reffig{fig:3}, we plot 
$\widehat{E_0}(\beta, 500, \alpha^*(500, \beta))$ 
as $\beta$ and $g$ are varied.
All the curves converge to zero and, in fact, can be smoothly
extrapolated to $\beta\to 0$. The resulting percentual
relative error $100|E_0-\widehat{E_0}|/|E_0|$ 
is very small, well below the permille level 
(see Tab.~(I) for numerical results with 4th order polynomial extrapolation).

For large coupling $g$,
the convergence is quite fast. The
one-parameter trial wave function is accurate because the ground state 
is dominated by states with low potential that are easily
selected by $|\Phi(\alpha)\rangle$. Relatively small $K$ 
are then already in the asymptotic regime.
For intermediate couplings, $g\sim 2.0$,
the convergence is again smooth, but less than linear.
For smaller couplings, a good convergence is observed and in fact
a precise wave function can obtained with $\alpha^*\simeq 0$.
The optimal $\alpha^*$ at $K=500$, $\beta = 0.1$ is shown in Tab.~(I).

\begin{figure}[h]
\centerline{\includegraphics*[width=4.75cm,angle=-90]{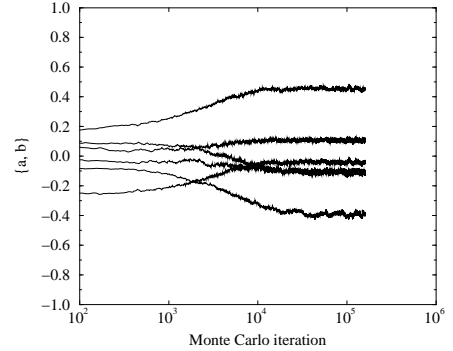}}
\caption{$L=10$, $N_f=2$, $g=2.0$.
MC evolution of the six parameters $\{a, b\}$.
From top to bottom, on the right of the plot, the parameters are 
$a_1$, $a_3$, $a_2$, $b_1$, $b_2$, $b_0$.}
\label{fig:4}
\end{figure}

\begin{figure}[h]
\centerline{\includegraphics*[width=4.75cm,angle=-90]{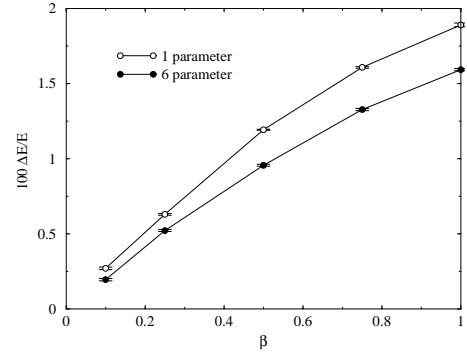}}
\caption{$L=10$, $N_f=2$, $g=2.0$, $K=500$. Improvement 
in the energy estimate with the 6-parameter 
trial wave function.}
\label{fig:5}
\end{figure}

For $g=2.0$, we explore a tentative 6-parameter trial
wave function. Denoting the two fermion flavors by
$\uparrow$, $\downarrow$, we use
$\langle {\bf n} | \Phi\rangle = e^{\sum_i F_i} \langle {\bf n} | g=0\rangle$
with 
$
F_i = \sum_{k=1}^{3} a_k (n_i^\uparrow n_{i+k}^\uparrow +
\uparrow\leftrightarrow \downarrow) + 
\sum_{k=0}^{2} b_k (n_i^\uparrow n_{i+k}^\downarrow +
\uparrow\leftrightarrow \downarrow) .
$
The MC automatic determination of the 6 parameters is shown
in~\reffig{fig:4}. The algorithm converges to definite
coefficients $\{a, b\}$, but the behavior of $\widehat{E_0}$
does not dramatically improve (\reffig{fig:5}). Nonetheless, some qualitative
remarks can be stressed, as the presence of long range correlations
between next to neighbor fermions with the same spin
and anti-correlations between fermions with opposite spin.

\begin{figure}[h]
\centerline{\includegraphics*[width=4.75cm,angle=-90]{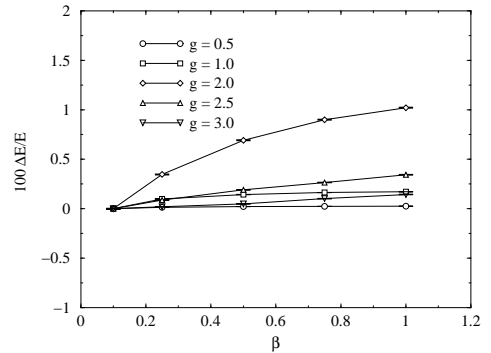}}
\caption{$L=10$, $N_f=6$, $K=500$. Relative percentual error on the
energy estimate obtained from data at large $K$ at several $\beta$.}
\label{fig:6}
\end{figure}

Since the Gross-Neveu model can be studied non perturbatively 
in the framework of the $1/N_f$ expansion, it is 
interesting to analyze the algorithm performance with a larger number
of flavors. In \reffig{fig:6}, we show the results for $N_f=6$. 
The exact value is beyond Lanczos diagonalization
and we choose to normalize errors
at the $\beta=0.1$, $K=500$ value. A comparison with~\reffig{fig:3}
reveals that the error as well as its $\beta$ dependence 
are rather reduced with respect to the previous $N_f=2$ case.

In summary, our data shows that a clever extrapolation in the 
algorithm free parameters $K$ and $\beta$
allows accurate results even 
with small walker ensembles. This is an important feature
for realistic large scale simulations aimed at reaching the
continuum limit.
Results with large $N_f$ suggest that the present algorithm 
can be a viable numerical technique for other
fermionic two-dimensional models where the $1/N_f$ expansion applies,
like the important case of models with dynamical supersymmetric breaking~\cite{susy}.
In principle, extensions to models with 
sign-problems are possible and, in fact, 
progress in the optimization issue has been recently proposed~\cite{sorella}
within the considered class of MC algorithms.

\vfill\end{multicols}
\widetext
\begin{table}
%\caption{Ground state energy per flavor $E_0/N_f$. ED = Exact
%Diagonalization, EX = polynomial extrapolation.}
%\begin{tabular}{l|cccc}
%$g$ & $N_f=2$ exact & $N_f=2$ extr. & 1000 $|\Delta E/E|$ & $N_f=6$ extr. \\
%\tableline
%0.5& 	-3.34904& 	-3.34908& 		0.012&	-3.27387\\
%1.0& 	-3.71687& 	-3.71689& 		0.005&	-3.40462\\
%2.0& 	-5.99265& 	-5.99285& 		0.03&	-5.70065\\
%2.5& 	-8.4526& 	-8.4524& 		0.02&	-8.26635\\
%3.0& 	-11.6949& 	-11.6927& 		0.2&	-11.5648
%\end{tabular}
\caption{$E_0/N_f$ for the $L=10$ model
with $N_f=2$ flavors. $\Delta E = E_0^{Lanczos}-E_0^{MC}$.}
\begin{tabular}{l|cccc}
$g$ & $\alpha^*(500, 0.1)$ & Exact Lanczos Diagonalization & Polynomial Extrapolation & 1000 $|\Delta E/E|$\\
\tableline
0.5& 	0.07638(1) &	-3.34904& 	-3.34908(5)& 		0.012\\
1.0& 	0.31347(5) &	-3.71687& 	-3.71689(5)& 		0.005\\
2.0& 	1.4044(3)  &	-5.99265& 	-5.9929(5)& 		0.03\\
2.5& 	2.0575(2)  &	-8.4526& 	-8.4524(3)& 		0.02\\
3.0& 	2.6198(2)  &	-11.6949& 	-11.6927(3)& 		0.2
\end{tabular}
\end{table} 

\begin{multicols}{2}
\narrowtext

\noindent
\noindent
\noindent
\noindent
\noindent
\noindent

\end{multicols}
\end{document}